# Neural-network models of memory consolidation and reconsolidation

**Peter Helfer (peter.helfer@mail.mcgill.ca)**
Department of Psychology, McGill University, 2001 McGill College Ave., 7th floor
Montreal, QC H3A 1G1 Canada

**Thomas R. Shultz (thomas.shultz@mcgill.ca)**
Department of Psychology and School of Computer Science, McGill University, 2001 McGill College Ave., 7th floor
Montreal, QC H3A 1G1 Canada

**Abstract**

In the mammalian brain newly acquired memories depend on the hippocampus for maintenance and recall, but over time these functions are taken over by the neocortex through a process called *systems consolidation*. However, reactivation of a consolidated memory can induce a brief period of temporary hippocampus-dependence, followed by return to hippocampus-independence. Here we present a computational model that uses simulation of recently described mechanisms of synaptic plasticity to account for findings from the systems consolidation/reconsolidation literature and to make predictions for future research.

## Introduction

The neural processes that transform memories from short-term to long-term storage are collectively known as *memory consolidation*. They include relatively rapid intra-cellular changes that stabilize synaptic potentiation (*synaptic consolidation*) as well as slower and larger-scale processes that reorganize and restructure memory traces across brain systems. In particular the latter include modifications that gradually make memories independent of the hippocampal formation in the medial temporal lobe (*systems consolidation*).

Retrieval of a consolidated memory can trigger a process in which it transiently becomes unstable, but subsequently restabilizes into the consolidated state. This is known as *reconsolidation*, and like consolidation it can be observed both at the synaptic and systems levels. For overviews of memory consolidation and reconsolidation research, see Dudai (2004), Nader & Einarsson (2010), Hardt et al. (2010).

Synaptic consolidation and reconsolidation have recently been shown to involve rapid changes in the proportions of different kinds of neurotransmitter receptors in the synapse (Clem & Huganir, 2010; Hong et al., 2013; Kessels & Malinow, 2009; Plant et al., 2006). Here we present a computational model that demonstrates that such receptor exchanges at the synaptic level can account for consolidation and reconsolidation at the systems level.

## Synaptic consolidation and reconsolidation

Neurons generate electric signals called action potentials (APs) that travel down nerve fibers toward synapses where connections are made with other neurons. When an action potential reaches a synapse, the presynaptic neuron releases neurotransmitter molecules that bind to receptors inserted in the postsynaptic cell membrane, thereby triggering activity in the postsynaptic neuron. The amount of activity that is generated by the arrival of an action potential is a measure of synaptic strength, and it depends both on the amount of transmitter released and on the number and types of receptors on the receiving side of the synapse (Kandel, Dudai, & Mayford, 2014).

The amino acid glutamate is the most abundant neurotransmitter in the vertebrate nervous system (Platt, 2007). There are several types of glutamate receptors, among which the AMPA receptors (AMPARs) are chiefly responsible for mediating excitatory synaptic transmission. Thus, the strength of a glutamatergic synapse depends strongly on the number of AMPA receptors inserted in the postsynaptic membrane.

When a neuron is stimulated strongly enough to make it fire (generate an AP), participating synapses can be strengthened by a process called long-term potentiation

(LTP), which is associated with an increase in the number of inserted AMPARs (Malenka & Bear, 2004). There are different forms of LTP. Moderately strong stimulation gives rise to early-phase LTP (E-LTP), which lasts for at most a few hours and is implicated in short-term memory. More intense stimulation can trigger the induction of late-phase LTP (L-LTP), which is believed to be an important mechanism for long-term memory (Abraham, 2003). The establishment of L-LTP is called synaptic consolidation.

Recent research has shown that different AMPAR subtypes are associated with the different phases of LTP: E-LTP induction is characterized by a rapid increase in the number of calcium-permeable AMPARs (CP-AMPARs), while the establishment of L-LTP requires insertion of calcium-impermeable CI-AMPARs (Clem & Huganir, 2010; Hong et al., 2013; Kessels & Malinow, 2009; Plant et al., 2006). The limited persistence of E-LTP reflects the degradation and/or removal from the synapse of CP-AMPARs, whereas the long, possibly unlimited, persistence of L-LTP has been hypothesized to be due to mechanisms that replenish and sequester CI-AMPARs at the synapse (Helfer & Shultz, 2018; Sacktor, 2011). Memory retrieval has been shown to trigger a transient reversal to the high CP-AMPAR/low CI-AMPAR state (Hong et al., 2013). The subsequent re-establishment of L-LTP is called synaptic reconsolidation. If synaptic reconsolidation is blocked by pharmaceutical means, then L-LTP is lost, and this has been shown to correlate with memory impairment (Nader & Hardt, 2009).

To summarize, moderate stimulation induces E-LTP, characterized by an increased number of CP-AMPARs which have a limited dwell time at the synapse. More intense stimulation triggers induction of L-LTP, which is associated with an increased number of CI-AMPARS and can persist for months or longer. Memory retrieval can cause a consolidated synapse to temporarily return to an unstable E-LTP-like state with high CP-AMPAR count and low CI-AMPAR count.

## Systems consolidation and reconsolidation

**Consolidation:** In the mammalian brain newly acquired memories depend on the hippocampus (HPC) for maintenance and recall, but over time these functions are taken over by the neocortex through a process called *systems consolidation* (Dudai, 2004). Lesion studies have shown that hippocampal involvement is required for systems consolidation to take place: hippocampal lesions impair new memories but not older ones (McClelland, McNaughton, & O'Reilly, 1995; Scoville & Milner, 1957). Different neocortical areas are important for different kinds of memories (Frankland & Bontempi, 2005); here we focus on conditioned fear memories in rodents, which have been shown to consolidate in the anterior cingulate cortex (ACC). Studies using pharmaceutical inactivation of hippocampus and/or ACC have shown that retrieval of a fear memory is hippocampus-dependent three days after acquisition, but not at 30 days. At this point it has instead become dependent on the ACC for retrieval (Einarsson, Pors, & Nader, 2015; Frankland, Bontempi, Talton, Kaczmarek, & Silva, 2004; Sierra et al., 2017).

**Reconsolidation:** Reactivation of a consolidated memory by presentation of a reminder stimulus can temporarily make retrieval ACC-independent again: six hours after retrieval, the memory is accessible even if the ACC is inactivated. HPC inactivation at this point also does not impair recall. However, simultaneous inactivation of both ACC and HPC blocks retrieval. Twenty-four hours after reactivation, retrieval has returned to being ACC-dependent and HPC-independent (Einarsson et al., 2015).

A hippocampal lesion performed within the first few hours after reactivation can cause permanent impairment or loss of the reactivated memory (Debiec, LeDoux, & Nader, 2002; Land, Bunsey, & Riccio, 2000; Winocur, Frankland, Sekeres, Fogel, & Moscovitch, 2009). This is in contrast with consolidated memories that have not been reactivated, or have been allowed to reconsolidate after reactivation. It thus appears that reactivation renders the ACC trace unstable and HPC involvement is needed for its restabilization.

In summary, retrieval of young (e.g. 3-day-old) fear memories requires the HPC but not the ACC. Over time a reversal takes place so that retrieval of 30-day-old memories requires the ACC but not the HPC. Reactivation of a consolidated memory temporarily returns it to ACC-independence for retrieval. Systems consolidation (establishment of an ACC trace) and systems reconsolidation (restabilization after reactivation-induced destabilization of the ACC trace) both require hippocampal activity.

## Model

The foregoing findings suggest the following models at the synaptic and systems level:

### Synaptic level

- Moderately intense stimulation induces E-LTP, which involves the rapid insertion of CP-AMPARs. Constitutive processes subsequently remove them within hours.
- L-LTP induction (synaptic consolidation) is a state change in a bistable mechanism (molecular switch), brought about by more intense stimulation. When in the ON state, this mechanism maintains a high count of CI-AMPARs in the synapse.
- Memory retrieval abruptly removes CI-AMPARs from the synapse and replaces them with CP-AMPARs, thus returning the synapse to an E-LTP-like state. The CI-AMPARs are subsequently restored. L-LTP is thus reestablished.

### Systems level

- Stimulus presentations trigger patterns of activation in multiple ensembles of neurons in the neocortex (NC). These active neurons in turn project onto and activate neurons in the hippocampus (HPC), where a memory trace is quickly created, providing linkages between the activated NC ensembles.

- Subsequently the HPC memory trace is spontaneously and repeatedly activated which causes stimulation of these same NC ensembles through nerve fibers projecting back from the HPC to the NC. Over time, Hebbian learning creates intra-neocortical connections, e.g. through the ACC, and strengthens them to a point where they can support retrieval of the memory without assistance of the hippocampus. This process is known as systems consolidation.
- Meanwhile, the HPC trace is gradually weakened by constitutive decay processes (Sachser et al., 2016).
- If the memory is reactivated, then the activity in the NC neural ensembles triggers recreation of the HPC linkage. Reactivation also destabilizes the ACC linkage.
- In the period following reactivation, the systems reconsolidation window, HPC stimulation of the now destabilized synapses of the intra-NC (ACC) linkage triggers the re-establishment of L-LTP in these synapses. The reactivated HPC trace decays rapidly, leading to a return to ACC-dependence in 24 hours or less.

## Computational modeling

Several artificial neural network (ANN) simulations of systems consolidation have been published (Alvarez and Squire (1994), McClelland (1995) and Murre (Meeter & Murre, 2005; Murre, 1996). These models all demonstrate how spontaneous activation of hippocampal traces can strengthen neocortical connections. However, due to the simplicity of the connections used in these models – the state of a synapse is modeled by a single variable, connection strength – they are not able to reproduce findings that involve variable synaptic stability, such as post-reactivation instability and reconsolidation. We here introduce a more elaborate connection model that allows our network to capture this wider range of empirical findings.

## Methods

### Network architecture

Like most artificial neural networks, ours consists of units and connections, where a unit is an analog of a biological neuron and a connection models a synapse.

**Topology:** We use a recurrent artificial neural network with four regions representing HPC, ACC and two sensory cortex areas, SC0 and SC1, to which stimuli are presented. Each region consists of 25 units. Each HPC unit is bidirectionally connected to each unit in SC0 and SC1, and similarly each ACC unit is connected to each SC0 and SC1 unit, see Figure 1.

**Units:** The units are bistable and stochastic; the probability that a unit will be active in the next time step is an asymmetric sigmoid function of net input,

$$P_j(net_j) = \frac{1}{1+e^{-\frac{net_j}{T}}} \quad (1)$$

where $net_j$, the net input to unit $j$, is the sum of the activity levels of units connected to unit $j$, weighted by inbound connection strengths.

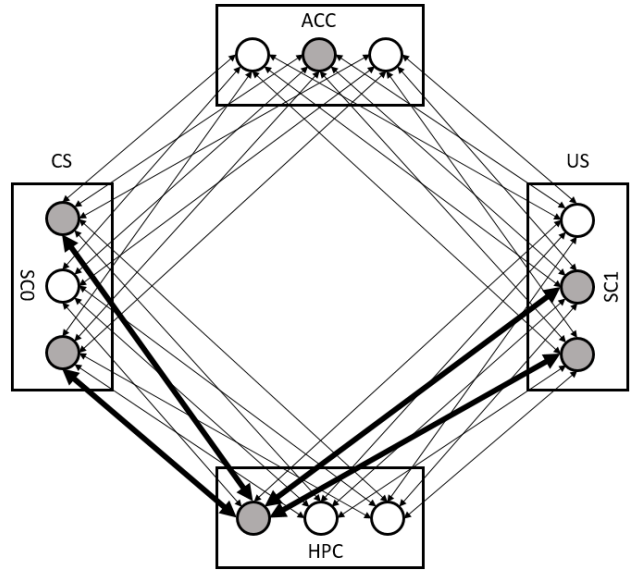

Figure 1: Network architecture. To reduce clutter only three units are shown in each region. Each double-arrow represents two independent connections, one in each direction between a pair of units. The diagram illustrates the state after initial acquisition: presentation of US and CS stimuli has activated some units in SC0 and SC1 (filled circles) and fast learning has created strong linkages (bold lines) through HPC. Linkages through ACC are still weak.

**Connections:** The connections are abstract models of glutamatergic synapses, characterized by four attributes: *capacity*, number of inserted CP-AMPARs and CI-AMPARs and a Boolean attribute *isPotentiated* that models the bistable nature of L-LTP. Learning is modeled by increasing the *capacity* attribute, allowing more AMPARs to be inserted. A connection's weight is equal to its total number of inserted AMPARs. The set of connections between two regions, e.g. from HPC to SC1, is referred to as a *tract*.

## Simulation

A simulation consists of a sequence of time steps. Various interventions may be scheduled for any time point during the simulation, and in addition several background processes execute at each time step. The scheduled event types are training, reactivation, HPC lesion, HPC inactivation and ACC inactivation. The background processes are consolidation, AMPAR trafficking and random depotentiation. In addition, a retrieval test can be executed at any time. The different interventions and background processes are described in the following.

**Learning rule:** The network learns activation patterns by increasing the capacity of connections between activated units:

$$c_{ij} = c_{ij} + \mu(c_{max} - c_{ij}) \quad (2)$$

where $c_{ij}$ is the capacity of the connection between units $i$ and $j$, $c_{max}$ is the maximum connection capacity and $\mu$ is a learning rate specific to the tract that the connection belongs to.

Capacity growth is accompanied by an increase of the number of CP-AMPARs so that the total AMPAR count equals the connection capacity. This models the rapid CP-AMPAR influx during E-LTP induction. In addition, probabilistic induction of L-LTP in a connection is simulated by turning on its *isPotentiated* attribute with a probability that depends on the stimulation strength.

Learning happens when stimuli are presented for training and at memory retrieval (reactivation), and also when patterns are spontaneously activated by the memory consolidation process. These mechanisms are described in the following.

**Training:** Training is simulated by activating subsets of units in SC0 and SC1 to represent an unconditioned stimulus, US, and a conditioned stimulus, CS, respectively. The network randomly selects and activates linkage units in HPC and ACC and then applies the learning rule to connections between active units. The learning rate is relatively high in HPC, allowing rapid creation of linkages strong enough to support recall. The ACC learning rate is lower, hence linkages through the ACC are not strong enough to support recall immediately after training.

**Retrieval:** To test recall of a trained pattern, the CS units are activated in SC0, and the network is cycled by repeated application of the activation function in all units. The activity pattern that the SC1 region then settles on may be compared to the associated US pattern to calculate a recall test score.

**Systems Consolidation:** At every simulation time step a random pattern is activated in HPC and then the entire network is cycled in the same manner as for recall test (but without stimulus presentation). Whatever pattern the network settles into is then reinforced by application of the learning rule. Because the network is more likely to settle into trained patterns than other random states, this will tend to strengthen CS-US linkages through the ACC, eventually making recall of trained patterns HPC-independent.

**AMPAR trafficking:** At each time step, the numbers of AMPARs in all connections are adjusted according to the following rules: The number of CP-AMPARs declines exponentially towards zero. If the connection's *isPotentiated* attribute is true, and the unit that the connection terminates on is receiving HPC input, then the number of CI-AMPARs grows asymptotically towards the number of available slots in the connection, i.e. *capacity* minus the number of CP-AMPARs, otherwise CI-AMPARS also decline exponentially.

**Depotentiation:** Potentiated connections are subject to random depotentiation. This happens with higher probability in HPC than in the neocortical regions, modeling the observed faster decline of hippocampal traces over time.

**Reactivation:** Reactivation is modeled as an unreinforced CS presentation, i.e. a cued retrieval. The CS pattern is activated in the SC0 region, the network is cycled, and when it settles the following processing takes place in all connections between pairs of active units:

- **AMPAR exchange**: the number of CI-AMPARs is reduced to a configured minimum, after which all available slots are filled with CP-AMPARs.
- **Depotentiate:** the *isPotentiated* flag is turned off.

This puts the ACC linkage in an unstable E-LTP-like state, simulating the observed post-reactivation instability. A set of HPC linkage units is then activated, as during initial acquisition, and a round of Hebbian learning takes place.

As noted in the introduction, the hippocampal engagement is much briefer after reactivation (less than 24h) than after initial training, when recall is HPC-dependent for at least three days. The mechanism underlying this faster disengagement is not known. One possibility is that memory reactivation triggers activation of a neuromodulatory factor that inhibits L-LTP induction in the HPC links. We chose to include such a factor in order to model the fast decay of HPC linkage after reactivation; other mechanisms are possible.

**Hippocampal lesion:** the HPC layer is disconnected from the simulation.

**Inactivation:** Inactivation of HPC or ACC is modeled by inhibiting activation of any units in the affected region.

## Results

Our model reproduces many findings reported in the systems consolidation/reconsolidation literature. A selection is described here; for a fuller treatment see Helfer & Shultz (in preparation).

1. HPC lesions produce memory deficit when performed in the consolidation and reconsolidation windows, but not otherwise. See Figure 2 and Figure 3. (The values in all diagrams are means of 100 simulation runs. The error bars show standard deviation.)

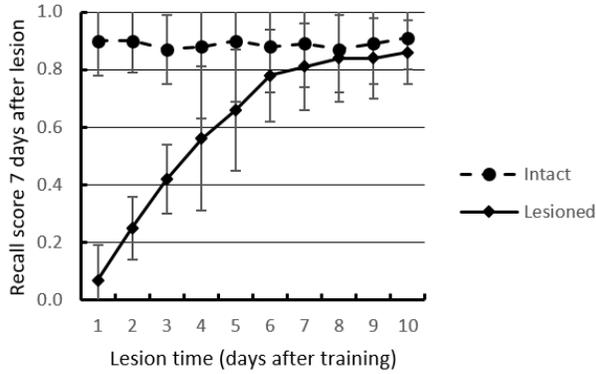

Figure 2: Consolidation window. Simulated HPC lesions produce severe impairment when performed shortly after training, but not later.

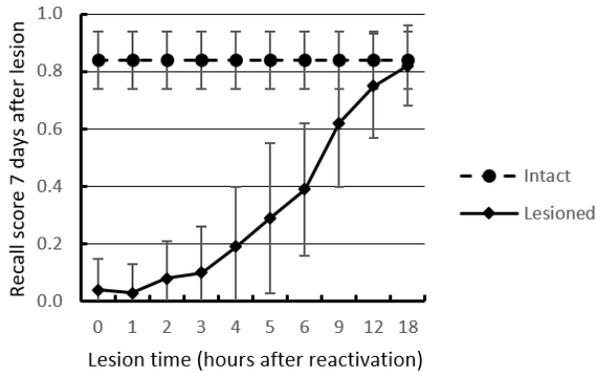

Figure 3: Reconsolidation window: Simulated HPC lesions produce severe impairment when performed shortly after reactivation, but not later.

2. Consolidation transforms memories from being HPC-dependent to being ACC-dependent for recall, see Figure 4.

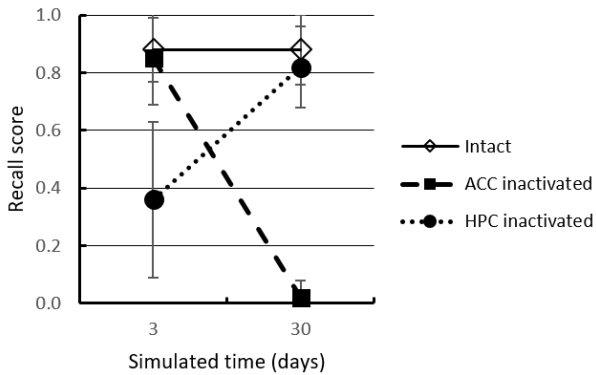

Figure 4: HPC/ACC-dependence for recall. Simulation of HPC inactivation impairs recall 3 days after training, but not at 30 days. Simulated ACC inactivation does not affect recall 3 days after training, but causes severe impairment at 30 days.

3. Reactivation creates a transient HPC linkage which temporarily returns a memory to ACC-independence, see Figure 5.

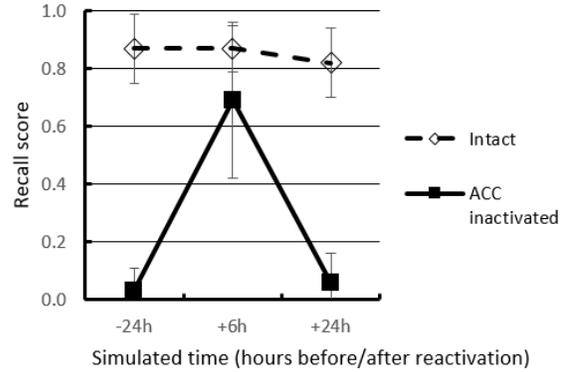

Figure 5: Temporary ACC-independence after reactivation. Before reactivation ACC inactivation impairs recall of a consolidated memory. Six hours after reactivation ACC inactivation produces no impairment. At 24h after reactivation, ACC dependence has returned.

## Discussion

We have presented an artificial neural network model of systems memory consolidation and reconsolidation that accounts for well-established findings from lesion studies as well as more recent results obtained using non-destructive inactivation methods. At the core of the model is a new connection design with variable stability arising from simulation of receptor exchanges that have been observed in glutamatergic synapses.

It is worth noting that although the term "reconsolidation" suggests a recapitulation of consolidation, our model reflects our view that the two processes are quite different. Whereas systems consolidation is a gradual strengthening of the neocortical synapses involved in a memory trace, systems reconsolidation is the restoration of L-LTP in such synapses following reactivation-induced destabilization. HPC lesion in the "consolidation window" (2 weeks or more following training) and in the "reconsolidation window" (six hours or less following reactivation) both result in memory deficits – but for different reasons: HPC lesions in the consolidation window terminate the spontaneous activations that drive strengthening of neocortical synapses, leaving a weak memory trace there. Loss of HPC in the reconsolidation window, in contrast, deprives neocortex of the HPC stimulation required to restore L-LTP in synapses that have been destabilized by reactivation-induced AMPAR exchange, resulting in a decay process similar to that observed in E-LTP.

The model thus predicts that if reactivation is prevented from triggering AMPA receptor exchange in the ACC, then HPC lesion in the reconsolidation window will not impair recall. This could be tested by infusing a drug like GluA2$_{3Y}$ into the ACC. GluA2$_{3Y}$ is a synthetic peptide that prevents endocytosis (removal) of CI-AMPARs from the synapse. In contrast, infusion of the same drug should not be able to prevent the recall-impairing effect of hippocampal lesion in the consolidation window.